\documentclass[12pt]{iopart}

\usepackage{indentfirst}

\begin{document}

\title[Algebraic derivation of Kramers-Pasternack relations]{Algebraic derivation of Kramers-Pasternack relations based on the Schr\"odinger factorization method}

\author{Tomasz Szymanski$^1$ and J. K. Freericks$^2$}

\address{$^1$Faculty of Physics and Astronomy, University of Wroclaw, 50-204 Wroclaw, Poland}
\address{$^2$Department of Physics, Georgetown University, Washington, DC 20057, USA}
\vspace{10pt}
\begin{indented}
\item[]\today
\end{indented}

\begin{abstract}
The Kramers-Pasternack relations are used to compute the moments of $r$ (both positive and negative) for all radial energy eigenfunctions of hydrogenic atoms. They consist of two algebraic recurrence relations, one for positive powers and one for negative. Most derivations employ the Feynman-Hellman theorem or a brute-force integration to determine the second inverse moment, which is needed to complete the recurrence relations for negative moments. In this work, we show both how to derive the recurrence relations algebraically and how to determine the second inverse moment algebraically, which removes the pedagogical confusion associated with differentiating the Hamiltonian with respect to the angular momentum quantum number $l$ in order to find the inverse second moment.
\end{abstract}

%
\vspace{2pc}
%
\submitto{European Journal of Physics}
%
%
%

\section{Introduction}

The expectation values of the powers of $r$ for the energy eigenfunctions of the Coulomb problem are often presented in quantum-mechanics textbooks. Sometimes only a few moments are stated without indicating that all can be computed or proving the ones stated (examples include Ballentine~\cite{ballentine}, Robinett~\cite{robinett} and Landau and Lifshitz~\cite{landau_lifshitz}). In other cases, they are computed by brute force in the position representation by using properties of Laguerre polynomials and integration by parts (Bransden and Joachain~\cite{bransden_joachain}, Commins~\cite{commins} and Messiah~\cite{messiah}). A complete treatment computes them via recurrence relations based on the hypervirial relation, which we will develop below. There turn out to be two different recurrence relations. The first relation, known as the first Pasternack relation or the Pasternack inversion relation, was discovered by Pasternack in the late 1930s~\cite{pasternack}. It is an identity quoted much less frequently in quantum textbooks than the second one. For this first one, Pasternack employed the generating function for the Laguerre polynomial published in the 1920s by Waller~\cite{waller}. Relying on the formulas obtained by Waller, Pasternack used the generalized hypergeometric series to construct the recurrence relation:
\begin{equation}  \label{Past1}
	\langle n,l |\hat{r}^{-m-2}|n,l \rangle = \left( \frac{2}{n a_0} \right)^{2m+1} \frac{\left( 2l - m \right)!}{\left( 2l + m +1 \right)!} \langle n,l | \hat{r}^{m-1} | n,l \rangle, 
\end{equation}
where $0\le m\le 2l$, $a_0$ is the Bohr radius and the state $|n,l\rangle$ is labeled by its principal quantum number $n$ and its total angular momentum $l$. We will discuss this in more detail below, but we note that when $m=0$ it relates the $1/r$ moment to the $1/r^2$ moment. We use that case as the base case for a proof by induction, which we give below; since it is the base case, one cannot use it as a substitute to directly calculating the inverse second moment. Interestingly, though Pasternack proved this result only for $0\le m\le 2l$, the formula seems to work for negative values of $m$ as well. This so-called Pasternack inversion relation was rediscovered (at least) twice: once by Bockasten~\cite{bockasten} in the 1970s and later by More~\cite{more} in the 1990s. It was also generalized by Blanchard~\cite{blanchard} for the off-diagonal matrix elements.

The second recurrence relation relates three consecutive moments to each other and is given by
\begin{eqnarray} \label{Past2}
	0 =& -\frac{2m}{n^2}\frac{1}{a_0^2} \langle n,l |\hat{r}^{m-1}|n,l \rangle + 2(2m-1)\frac{1}{a_0} \langle n,l |\hat{r}^{m-2}|n,l \rangle \\
	     &- \frac{1}{2}(m-1)\left ((2l+1)^2-(m-1)^2\right ) \langle n,l |\hat{r}^{m-3}|n,l \rangle;
\end{eqnarray}
it is known as the second Pasternack relation, the Kramers-Pasternack relation or, sometimes, the Kramers relation. It was independently developed in the  1930's by Pasternack~\cite{pasternack} and by Kramers~\cite{kramers} (in his textbook, which was not translated into English until the late 1950s). While Pasternack obtained this recurrence relation by further manipulations of Waller's result with the generalized hypergeometric series, Kramers' method relied on manipulations of the radial equation. The radial equation is first multiplied by an expression that is closely related to terms used in the hypervirial theorem (but is unmotivated in the text). Then, after several integrations by parts, he obtains the second Pasternack formula. A different method of derivation was presented by Epstein and Epstein~\cite{epstein_epstein} in the early 1960s. Their method is purely algebraic and relies on the use of the hypervirial theorem. It is likely that students accustomed to operator methods will find this method clearer and easier to follow. It has also been adopted in textbooks, such as B\"ohm~\cite{bohm} or De Lange and Raab~\cite{delange_raab}.

While the inversion relation is often ignored in quantum mechanics textbooks, the Kramers-Pasternack relation is often presented and used in calculations (usually for the perturbation theory of the fine structure of hydrogen). Sometimes it is simply stated without proof (some examples include Banks~\cite{banks}, Basdevant and Dalibard~\cite{basdevant_dalibard} and Zettili~\cite{zettili}). Other times its proof is left as an exercise, usually with the method developed by Kramers (some examples include Fitts~\cite{fitts}, Griffiths~\cite{griffiths}, Liboff~\cite{liboff}, Messiah~\cite{messiah}, Nolting~\cite{nolting} and Schwabl~\cite{schwabl}). The textbooks by Adams~\cite{adams}, B\"ohm~\cite{bohm} and De Lange and Raab~\cite{delange_raab} present algebraic proofs based on the hypervirial theorem. Interestingly, Shankar~\cite{shankar}, Sakurai~\cite{sakurai} and Townsend~\cite{townsend} obtain $\langle n,l |\hat{r}^{-2}|n,l \rangle$ using clever perturbative tricks and then establish the relation between $\langle n,l |\hat{r}^{-2}|n,l \rangle$ and $\langle n,l |\hat{r}^{-3}|n,l \rangle$ via the hypervirial theorem.

The second recurrence relation is a two-term recurrence relation. So one might have thought that by knowing both the zeroth moment (which is one from normalization) and the first inverse moment (which is easy  to find from the virial theorem) that we obtain all the rest. But there is a problem with this approach. Substituting $m=1$ into Eq.~(\ref{Past2}) causes one of the coefficients to be zero and yields 
\begin{equation} \label{power1}
	\langle n,l |\hat{r}^{-1}|n,l \rangle = \frac{1}{a_0 n^2}.
\end{equation}

This stymies the determination of  $\langle n,l |\hat{r}^{-2}|n,l \rangle$ from the original recurrence relation (we can find all positive moments, but not the negative ones). Hence, to find the negative moments requires an independent determination of the $1/r^2$ moment. Many textbooks prefer a brute force method to calculate $\langle n,l|1/\hat r^2|n,l\rangle$ by integrating Laguerre polynomials (Basdevant and Dalibard~\cite{basdevant_dalibard}, B\"ohm~\cite{bohm}, Liboff~\cite{liboff}, Schwabl~\cite{schwabl} and Zettili~\cite{zettili}). While the integrals  are straightforward, the computation is tedious and cumbersome. It also requires proficiency in working with Laguerre polynomials. This may be the reason why some authors advocate to use the Feynman-Hellman theorem instead (Adams~\cite{adams}, Banks~\cite{banks},
Griffiths~\cite{griffiths} and Fitts~\cite{fitts}). This method can be shown to be mathematically rigorous~\cite{sanchez_del_rio,fernandez_castro}, but it is likely to cause confusion amongst students. This is because it requires the replacement of a discrete quantum number $l$ (which determines the total angular momentum) by a continuous variable that is then differentiated. In solving the energy eigenstates of the hydrogen atom, students are carefully instructed that the angular momentum quantum number $l$ is discrete. Hence, students are likely to have difficulty following this derivation, without significant additional instruction that demonstrates why this manipulation is allowed.     

To address this problem, we propose a  purely algebraic method of computing $\langle n,l |\hat{r}^{-2}|n,l \rangle$. It is based on the Schr\"odinger factorization method and should be accessible to a wide group of students. It also does not require one to assume that $l$ becomes continuous; on the contrary, it uses and embraces its discreteness. 

%
%
%
%

\section{Algebraic Derivation of the Kramers-Pasternack Relations}


We first employ the Schr\"odinger factorization method~\cite{schrodinger,green,ohanian} to compute the energy eigenfunctions of the hydrogen atom. The traditional way computes the energy eigenfunctions from a sequence of raising operators acting on an auxiliary Hamiltonian ground state. But, it turns out that because the Coulomb problem has extra symmetry~\cite{pauli,fock}, one can use a simpler methodology to find the energy eigenfunctions all at once~\cite{rushka} because the auxiliary Hamiltonians correspond to physical Hamiltonians with different angular momentum.

We start from these Hamiltonians in each angular momentum sector, which are determined after separation of variables, and produce a different radial equation for each $l$. The first step, is to factorize each Hamiltonian, using ladder operators $\hat{B}_l=\frac{1}{\sqrt{2\mu}}(\hat{p}_r-i\frac{\hbar}{a_0}W(\hat r))$, where $\mu$ is the reduced mass, $\hat{p}_r$ is the radial momentum and $W(\hat r)$ is called the superpotential. The radial momentum satisfies $[\hat r,\hat p_r]=i\hbar$ and $[\hat p_r,\hat r^m]=-i\hbar m \hat r^{m-1}$. To determine the ladder operator, we take into account the condition the superpotential must satisfy, which is that $\lim_{r\to 0}W(r)=\infty$ and $\lim_{r\to\infty}W(r)>0$. A simple calculation then shows that
\begin{equation} \label{Bl}
	\hat{B}_l = \frac{1}{\sqrt{2\mu}} \left\{ \hat{p}_r - i\hbar \left( \frac{1}{\left( l+1 \right) a_0} - \frac{\left( l+1 \right)}{\hat{r}} \right) \right\},
\end{equation}
and
\begin{equation} \label{Hl}
	\hat{\mathcal{H}}_l = \frac{\hat{p}_r^2}{2\mu} + \frac{\hbar^2 l\left(l+1\right)}{2\mu\hat{r}^2} - \frac{e^2}{\hat{r}} = \hat{B}_l^{\dagger} \hat{B}_l + E_l.
\end{equation}
Here, $\hat B_l^\dagger$ is the Hermitian conjugate of $\hat{B}_l$, $e$ is the magnitude of the charge of the electron and the proton, $a_0 = \hbar^2/\mu e^2$ and $E_l = - \frac{e^2}{2 \left( l+1 \right)^2 a_0}$. Note that the standard notation for the energy of hydrogen uses the principal quantum number. Be careful that we  instead use the maximal angular momentum for the label (so that $n=l+1$).

We denote the eigenstate of $\hat{\mathcal{H}}_l$, corresponding to eigenvalue $E_{l{=}n-1}$ as $|n{=}l+1, l\rangle$ or, equivalently, as $|n,l{=}n-1\rangle$. Using the fact that $\hat{B}_l^{\dagger} \hat{B}_l$ is a positive semidefinite operator, its minimal eigenvalue is 0, which occurs when $\hat{B}_{n-1}|n,n-1 \rangle = 0$, which is called the subsidiary condition. In this case, the ground-state energy of $\hat{\mathcal{H}}_l$ $(\hat{\mathcal{H}}_{n-1})$ is $E_l$ ($E_{n-1}$). Note that we assume that the state $|n,n-1\rangle$ is normalized, so that $\langle n,n-1|n,n-1\rangle=1$.

In order to find the other degenerate energy eigenstates, we need to determine the so-called intertwining relation. Using the product of the ladder-operators in the opposite order, we find that
\begin{equation} \label{wrong_order}
	\hat{B}_l \hat{B}_l^{\dagger} = \frac{\hat{p}_r^2}{2\mu} + \frac{\hbar^2 \left(l+1\right)\left(l+2\right)}{2\mu\hat{r}^2} - \frac{e^2}{\hat{r}} - E_l = \hat{\cal{H}}_{l+1} - E_l.
\end{equation}
This result immediately establishes the so-called intertwining relation
\begin{equation} \label{inter}
	\hat{\cal{H}}_l \hat{B}_l^{\dagger} = \hat{B}_l^{\dagger} \hat{\cal{H}}_{l+1}.
\end{equation}
With the aid of Eq.~(\ref{inter}), we construct all of the energy eigenstates with energy $E_{n-1}$ (these are excited energy eigenstates of each $\hat{\mathcal H}_l$ for $0\le l<n-1$). We observe that the (unnormalized) states
\begin{equation} \label{eigenst}
	|n,l \rangle = \hat{B}_l^{\dagger} \hat{B}_{l+1}^{\dagger} \ldots \hat{B}_{n-3}^{\dagger} \hat{B}_{n-2}^{\dagger} |n,n-1 \rangle,
\end{equation}
where  $0 \leq l < n-1$, are each energy eigenstates of the corresponding Hamiltonian $\hat{\mathcal{H}}_l$ with energy $E_{n-1}$; the states will be normalized below. Using Eq.~(\ref{inter}), one finds that
\begin{eqnarray}
	\hat{\cal{H}}_l |n,l \rangle &= \hat{\cal{H}}_l \hat{B}_l^{\dagger} \hat{B}_{l+1}^{\dagger} \cdots \hat{B}_{n-3}^{\dagger} \hat{B}_{n-2}^{\dagger} |n,n-1 \rangle \nonumber\\
	&=\hat{B}_l^{\dagger}\hat{\cal{H}}_{l+1}\hat{B}_{l+1}^{\dagger} \cdots \hat{B}_{n-3}^{\dagger} \hat{B}_{n-2}^{\dagger} |n,n-1 \rangle \nonumber\\
	&=  \hat{B}_l^{\dagger} \hat{B}_{l+1}^{\dagger} \cdots \hat{B}_{n-3}^{\dagger} \hat{B}_{n-2}^{\dagger} \hat{\cal{H}}_{n-1} |n,n-1 \rangle \nonumber\\
	&=\hat{B}_l^{\dagger} \hat{B}_{l+1}^{\dagger} \cdots \hat{B}_{n-3}^{\dagger} \hat{B}_{n-2}^{\dagger} E_{n-1} |n,n-1 \rangle \nonumber\\
								   &= E_{n-1} |n,l \rangle.
\end{eqnarray}
Note that we cannot extend these eigenstates beyond $l=0$ because $\hat{B}_{-1}^\dagger$ is not well-defined. This result implies that the states $|n,l\rangle$ all have energy $E_{n-1}$ for $0\le l\le n-1$.

Now we determine the normalization constant $C_{nl}$, which we multiply the unnormalized eigenstates $|n,l\rangle$ by to make them normalized. Computing the norm, then yields
\begin{equation}
	1 =  |C_{nl}|^2 \langle n,n-1| \hat{B}_{n-2} \cdots \hat{B}_l \hat{B}_l^{\dagger} \cdots \hat{B}_{n-2}^{\dagger} |n,n-1 \rangle .
\end{equation}
We use Eq.~(\ref{wrong_order}) to convert the innermost product $\hat{B}_l\hat{B}^\dagger_l$ to $\hat{\mathcal{H}}_{l+1}-E_l$. Using the intertwining relation moves it to the right, where it becomes $\hat{\mathcal{H}}_{n-1}-E_l$, which becomes $E_{n-1}-E_l$ after operating the Hamiltonian onto the state $|n,n-1\rangle$. Repeating until all operators are removed, yields
\begin{equation}
    1=|C_{nl}|^2\prod_{i=l}^{n-2}(E_{n-1}-E_i).
\end{equation}
Plugging in the value of the energy then gives
\begin{equation}
    C_{nl}=\left (\frac{2a_0n^2}{e^2}\right )^{\frac{n-l-1}{2}}\frac{[(n-1)!]}{l!}\sqrt{\frac{(n+l)!}{(2n-1)!(n-l-1)!}}.
\end{equation}
We now absorb the normalization constant into the definition of the $|n,l\rangle$ states and work with normalized states only for the remainder of this work.

We are now ready to derive the Kramers-Pasternack identity algebraically. We employ the hypervirial theorem~\cite{epstein_epstein,adams,bohm} which, in our case, takes the form:
\begin{eqnarray} \label{hyper}
	\langle n,l| [\hat{\mathcal O}, \hat{\mathcal H}_l ] |n,l \rangle &= \langle n,l|\left (\hat {\mathcal O}\hat{\mathcal H}_l-\hat{\mathcal H}_l\hat {\mathcal O}\right )|n,l\rangle\nonumber\\
	&=E_{n-1}\langle n,l|(\hat {\mathcal O}-\hat {\mathcal O})|n,l\rangle=0,
\end{eqnarray}
because $\hat{\mathcal H}_l|n,l\rangle=E_{n-1}|n,l\rangle$. We require the states $|n,l\rangle$ and $\langle n,l|$ to be  elements of the domains of $\hat{\mathcal H}_l$  and of $\hat{\mathcal O}$ for the hypervirial theorem to be true. It is only if this vector (and its dual) is in both domains that we can actually evaluate the matrix elements in the hypervirial theorem.

%
%
 
In the ordinary virial theorem, we take $\hat {\mathcal O}=\hat r\hat p_r+\hat p_r\hat r$, but for the hypervirial theorem, we take $\hat{\mathcal O} = \hat{r}^m \hat{p}_r + \hat{p}_r \hat{r}^m$, and we then compute the commutator
\begin{eqnarray} \label{OHl}
	[ \hat{\mathcal O}, \hat{\mathcal H}_l ] &= \left[ \hat{r}^m \hat{p}_r + \hat{p}_r \hat{r}^m, \frac{\hat{p}_r^2}{2\mu} + \frac{\hbar^2 l\left(l+1\right)}{2\mu\hat{r}^2} - \frac{e^2}{\hat{r}} \right] \nonumber\\
						    &= \frac{1}{2\mu} [ \hat{r}^m \hat{p}_r + \hat{p}_r \hat{r}^m, \hat{p}_r^2 ] + \frac{\hbar^2 l\left(l+1\right)}{2\mu} \left [ \hat{r}^m \hat{p}_r + \hat{p}_r \hat{r}^m, \frac{1}{\hat{r}^{2}} \right ] \nonumber \\ 
						    &\quad\,-  \left [ \hat{r}^m \hat{p}_r + \hat{p}_r \hat{r}^m, \frac{e^2}{\hat{r}} \right ] \nonumber \\
						    &= \frac{i \hbar m}{2\mu} \left( \hat{r}^{m-1} \hat{p}_r^2 + 2 \hat{p}_r \hat{r}^{m-1} \hat{p}_r + \hat{p}_r^2 \hat{r}^{m-1} \right) \nonumber\\
						    & +\frac{i\hbar}{2\mu} 4\hbar^2 l(l+1) \hat{r}^{m-3} - 2 i\hbar e^2 \hat{r}^{m-2}.
\end{eqnarray}
Next, we rearrange the $2\hat p_r \hat r^{m-1}\hat p_r$ term by moving all $\hat p_r$ operators to the left for one term and to the right for the other. This gives
\begin{eqnarray}
	 2 \hat{p}_r \hat{r}^{m-1} \hat{p}_r &= \hat{p}_r \left( \hat{p}_r \hat{r}^{m-1} +  [ \hat{r}^{m-1}, \hat{p}_r ] \right) + \left( \hat{r}^{m-1} \hat{p}_r - [ \hat{r}^{m-1}, \hat{p}_r ] \right) \hat{p}_r \nonumber\\
	 &=  \hat{p}_r^2 \hat{r}^{m-1} + \hat{r}^{m-1} \hat{p}_r^2 + \hbar^2\left(m-1\right)\left(m-2\right)\hat{r}^{m-3}.\nonumber
\end{eqnarray}
We substitute this into Eq.~(\ref{OHl}) to find
\begin{eqnarray}
	[ \hat{\mathcal O}, \hat{\mathcal H}_l ] = i\hbar \Bigg\{& 2m\frac{\hat{p}_r^2}{2\mu}\hat{r}^{m-1} + 2m\hat{r}^{m-1}\frac{\hat{p}_r^2}{2\mu}  \\
		 	 							      &+ \frac{\hbar^2}{2\mu}\Big(m\left(m-1\right)\left(m-2\right) + 4l\left(l+1\right) \Big)\hat{r}^{m-3} - 2e^2\hat{r}^{m-2} \Bigg\}. \nonumber
\end{eqnarray}
Now, we recognize that we can substitute in Eq.~(\ref{Hl}) for $\hat{\mathcal H}_l$ twice, which gives 
\begin{eqnarray} \label{OHl_res}
	[ \hat{\mathcal O}, \hat{\mathcal H}_l ] &= i\hbar \Bigg\{ 2m\hat{\cal{H}}_l\hat{r}^{m-1} +  2m\hat{r}^{m-1}\hat{\cal{H}}_l  \\
						    				      &- \frac{\hbar^2}{2\mu}(m-1)\left(\left(2l+1\right)^2 - \left(m-1\right)^2 \right)\hat{r}^{m-3} + 2\left(2m-1\right) e^2 \hat{r}^{m-2} \Bigg\}.\nonumber
\end{eqnarray}
Using $\hat{\cal{H}}_l |n,l \rangle = E_{n-1} |n,l \rangle$ in the hypervirial theorem finally establishes that
\begin{eqnarray}
	0 = i\hbar \Bigg\{& -2m\frac{e^2}{a_0 n^2} \langle n,l| \hat{r}^{m-1} | n,l \rangle + 2\left( 2m-1 \right)e^2 \langle n,l| \hat{r}^{m-2} | n,l \rangle \nonumber \\
					    & -\frac{\hbar^2}{2\mu} \left(m-1\right)\left(\left(2l+1\right)^2 - \left(m-1\right)^2 \right) \langle n,l| \hat{r}^{m-3} | n,l \rangle \Bigg\}.
\end{eqnarray}
Using $a_0 = \hbar^2/\mu e^2$, we obtain Eq.~(\ref{Past2}), which is the famous Kramers-Pasternack relation (or the second Pasternack relation).

Armed with this formula, we can immediately determine the expectation values for $m > 0$. First note that the $m=0$ moment satisfies $\langle n,l|n,l\rangle=1$, because it is a normalized state. Setting $m = 1$ in Eq.~(\ref{Past2}), we find 
\begin{equation} \label{1/r}
	\langle n,l |\hat{r}^{-1}|n,l \rangle = \frac{1}{a_0n^2},
\end{equation}
which quite often is presented in textbooks (see for example, ~\cite{shankar} or ~\cite{sakurai}) as an immediate consequence of the virial theorem. Next we set $m = 2$ in Eq.~(\ref{Past2}) to obtain (using the previous result in Eq.~(\ref{1/r}))
\begin{equation}
	\langle n,l |\hat{r}|n,l \rangle = \frac{a_0}{2} \left( 3n^2 - l \left( l+1 \right) \right).
\end{equation}
Setting $m = 3$ in Eq.~(\ref{Past2}) yields
\begin{equation}
	\langle n,l |\hat{r}^2|n,l \rangle = \frac{a_0^2 n^2}{2} \left( 5n^2 + 1 - 3l \left( l+1 \right) \right).
\end{equation}

Though the formulas we obtain for larger $m$ become more and more complicated, we can proceed this way as far as we wish. We encounter a difficulty, however, when we try to determine the negative moments with Eq.~(\ref{Past2}). Setting $m = 0$ yields
\begin{equation}
	\langle n,l |\hat{r}^{-3}|n,l \rangle = \frac{1}{a_0 l \left( l+1 \right) }\langle n,l |\hat{r}^{-2}|n,l \rangle,\label{eq:3_to_2}
\end{equation}
because the coefficient of the $-1$ moment is zero. Hence, we need the inverse square moment to continue the recurrence relation for all subsequent inverse moments. This is a problem that is usually treated by brute force integration or by using the Feynman-Hellman theorem (and differentiating with respect to $l$).


We propose a new method for dealing with this problem based on the factorization framework outlined above. We start by observing that Eq.~(\ref{Hl}) implies that
\begin{equation}
	\Big\langle n,l\Big| (\hat{\cal{H}}_l - E_l) \frac{1}{\hat{r}^{2}}\Big|n,l \Big\rangle = \Big\langle n,l\Big| \hat{B}_l^{\dagger} \hat{B}_l \frac{1}{\hat{r}^{2}} \Big|n,l \Big\rangle .
\end{equation}
Operating the Hamiltonian to the left and dividing by the difference of energies, we then find an expression for the inverse second moment given by
\begin{equation}
	\Big\langle n,l\Big| \frac{1}{\hat{r}^{2}} \Big|n,l \Big\rangle = \frac{1}{E_{n-1} - E_l} \Big\langle n,l\Big| \hat{B}_l^{\dagger} \hat{B}_l \frac{1}{\hat{r}^{2}} \Big|n,l \Big\rangle .\label{st_point}
\end{equation}
The strategy is to move the $\hat{B}_l$ factor past the $\hat{r}^{-2}$ term so it can meet a $\hat{B}_l^{\dagger}$ operator that is in the operator expression for the $|n,l\rangle$ state in terms of the $|n,l{=}n-1\rangle$ state. This then allows us to use Eq.~(\ref{inter}) to replace the $\hat B_l\hat B_l^\dagger$ term in terms of the Hamiltonian for $l+1$; which can be moved to the right (or the left) to act against $|n,l{=}n-1\rangle$ (or $| n,n-1\rangle$) due to the intertwining relation.
This algebra is straightforward:
\begin{eqnarray}
	\Big\langle n,l\Big| \frac{1}{\hat{r}^{2}} \Big|n,l \Big\rangle &= \frac{1}{E_{n-1} - E_l} \Big\langle n,l\Big|\hat{B}_l^{\dagger} \, \frac{1}{\hat{r}^{2}} \,  \hat{B}_l  \Big|n,l\Big\rangle \nonumber\\
																		&+ \frac{1}{E_{n-1} - E_l} \Big\langle n,l\Big| \hat{B}_l^{\dagger}  \left[ \hat{B}_l \, , \frac{1}{\hat{r}^{2}}\right] \,  \Big|n,l \Big\rangle \\
																		&= \frac{1}{\left(E_{n-1} - E_l\right)^2} \Big\langle n,l+1\Big|\hat{B}_l \hat{B}_l^{\dagger} \, \frac{1}{\hat{r}^{2}} \, \hat{B}_l \hat{B}_l^{\dagger}  \Big|n,l+1 \Big\rangle \label{3rd_line} \nonumber\\ 
																		&+ \frac{1}{E_{n-1} - E_l} \frac{2i\hbar}{\sqrt{2\mu}} \Big\langle n,l\Big| \hat{B}_l^{\dagger} \, \frac{1}{\hat{r}^{3}} \Big|n,l \Big\rangle, 
\end{eqnarray}
where we used the fact that the normalized states satisfy $|n,l \rangle = \frac{1}{\sqrt{E_{n-1} - E_l}}\hat B_l^\dagger|n,l+1 \rangle$  and the commutator is easily evaluated to be $[\hat{B}_l^{\dagger}, \hat{r}^{-2}] = \frac{2i\hbar}{\sqrt{2\mu}} \hat{r}^{-3}$.
Next, we employ the intertwining relation from Eq.~(\ref{inter}) on the first term in Eq.~(\ref{3rd_line}) to move the Hamiltonian factors to the right, increasing the index by one with each step, until they reach the state  $|n,n-1\rangle$ on the right (and similarly on the left)
\begin{eqnarray}
	&\frac{1}{\left( E_{n-1} - E_l \right)^2} \Big\langle n,l+1\Big|\hat{B}_l \hat{B}_l^{\dagger}  \frac{1}{\hat{r}^{2}} \,  \hat{B}_l \hat{B}_l^{\dagger} \Big|n,l+1\Big\rangle \nonumber\\
	&=\frac{1}{\left( E_{n-1} - E_l \right)^2} \Big\langle n,l+1\Big|\left( \hat{\cal{H}}_{l+1} - E_l \right) \frac{1}{\hat{r}^{2}} \left( \hat{\cal{H}}_{l+1} - E_l \right) \Big|n,l+1\Big\rangle \\ 
	&=\Big\langle n,l+1\Big|\frac{1}{\hat{r}^{2}}\Big|n,l+1\Big\rangle. \label{long1}
\end{eqnarray}
Next we write $\hat{B}_l^{\dagger}$ out explicitly in terms of the momentum and position operators in the second term 
\begin{eqnarray}
	\Big\langle n,l\Big| \hat{B}_l^{\dagger} \frac{1}{\hat{r}^{3}} \Big|n,l \Big\rangle &= \frac{1}{\sqrt{2\mu}} \Big\langle n,l\Big| \Bigg\{ \hat{p}_r + i\hbar \Bigg( \frac{1}{(l+1)a_0} - \frac{l+1}{\hat{r}} \Bigg) \Bigg\} \frac{1}{\hat{r}^{3}} \Bigg|n,l \Big\rangle \nonumber\\
																								&= \frac{1}{\sqrt{2\mu}} \Big\langle n,l\Big| \hat{p}_{r} \frac{1}{\hat{r}^{3}} \Big|n,l \Big\rangle  \label{prr-3} + \frac{i\hbar}{\sqrt{2\mu}(l+1)a_0} \Big\langle n,l\Big| \frac{1}{\hat{r}^{3}} \Big|n,l \Big\rangle \label{r-3} \nonumber\\
																								&- \frac{i\hbar(l+1)}{\sqrt{2\mu}} \Big\langle n,l\Big| \frac{1}{\hat{r}^{4}} \Big|n,l \Big\rangle \label{r-4}
\end{eqnarray}
Two of these terms involve more negative moments. We focus first on the remaining term with the radial momentum. We apply the hypervirial theorem one more time in the form
\begin{equation} \label{hyper_r-2}
	\Big\langle n,l\Big|\left[\hat{\cal{H}}_l, \frac{1}{\hat{r}^{2}}\right]\Big|n,l \Big\rangle = 0.
\end{equation}
The commutator can be evaluated immediately
\begin{equation}
	\left[\hat{\cal{H}}_l, \frac{1}{\hat{r}^{2}} \right] = \frac{2i\hbar}{2\mu} \left( \hat{p}_r \frac{1}{\hat{r}^3} + \frac{1}{\hat{r}^3}\hat{p}_r  \right) = \frac{i\hbar}{\mu} \left( 2\hat{p}_r \frac{1}{\hat{r}^3} -\frac{3i\hbar}{\hat{r}^4} \right), \label{Hlr-2}
\end{equation}
after moving the momentum operators to the left.
Substituting into the hypervirial relation, we obtain
\begin{equation} \label{prr-3_res}
	\Big\langle n,l\Big| \hat{p}_{r} \frac{1}{\hat{r}^{3}} \Big|n,l \Big\rangle = \frac{3i\hbar}{2} \Big\langle n,l\Big|\frac{1}{\hat{r}^{4}} \Big|n,l \Big\rangle.
\end{equation}
This is then substituted into the right-hand side of Eq.~(\ref{prr-3}), which involves a sum over the inverse third and fourth moments. We now can relate the inverse third and inverse fourth moments to the inverse second moment. We start from the Kramers-Pasternack formula with $m=-1$:
\begin{equation} 
	0 = \frac{2}{n^2 a_0^2}\Big\langle n,l \Big|\frac{1}{\hat{r}^{2}}\Big|n,l \Big\rangle - \frac{6}{a_0}\Big\langle n,l \Big|\frac{1}{\hat{r}^{3}}\Big|n,l \Big\rangle + \left( \left(2l+1\right)^2-4\right)\Big\langle n,l \Big|\frac{1}{\hat{r}^{4}}\Big|n,l \Big\rangle.
\end{equation}
This relates the inverse fourth moment to a sum of the inverse third and inverse second moments. We use this to remove the inverse fourth moment from the right hand side of Eq.~(\ref{r-3}). Then we use Eq.~(\ref{eq:3_to_2}) to remove the inverse third moment. After some significant algebra, we find
\begin{equation}
	\Big\langle n,l\Big|\frac{1}{\hat{r}^{2}}\Big|n,l\Big\rangle = \Big\langle n,l+1\Big|\frac{1}{\hat{r}^{2}}\Big|n,l+1\Big\rangle + \frac{2}{2l+3}\Big\langle n,l\Big|\frac{1}{\hat{r}^{2}}\Big|n,l\Big\rangle,
\end{equation}
which can be rearranged to
\begin{equation}
	\Big\langle n,l\Big|\frac{1}{\hat{r}^{2}}\Big|n,l\Big\rangle = \frac{l+\frac{3}{2}}{l+\frac{1}{2}}\Big\langle n,l+1\Big|\frac{1}{\hat{r}^{2}}\Big|n,l+1\Big\rangle.
\end{equation}
Repeating this procedure $n-l-2$ times, we have
\begin{equation} \label{r-2n}
	\Big\langle n,l\Big|\frac{1}{\hat{r}^{2}}\Big|n,l\Big\rangle = \frac{n-\frac{1}{2}}{l+\frac{1}{2}}\Big\langle n,n-1\Big|\frac{1}{\hat{r}^{2}}\Big|n,n-1\Big\rangle.
\end{equation}
We next use the subsidiary condition $\hat B_{n-1}|n,n-1\rangle=0$ to determine the right-hand side. We observe that in the expression $\Big\langle n,n-1\Big|\left(\hat{B}_{n-1}^{\dagger} - \hat{B}_{n-1} \right)^2\Big|n,n-1\Big\rangle$ only one term survives
\begin{equation}
	\Big\langle n,n-1\Big|\left(\hat{B}_{n-1}^{\dagger} - \hat{B}_{n-1} \right)^2\Big|n,n-1\Big\rangle = -\Big\langle n,n-1\Big| \hat{B}_{n-1}\hat{B}_{n-1}^{\dagger} \Big|n,n-1\Big\rangle.
\end{equation}
Using the explicit forms for $\hat{B}_{n-1}^{\dagger}$ and $\hat{B}_{n-1}$, we find that the square of their difference involves only the zeroth and first and second inverse powers of $\hat r$, because the radial momentum terms cancel. In particular, we find
\begin{equation}
	\left(\hat{B}_{n-1}^{\dagger} - \hat{B}_{n-1} \right)^2 = -\frac{2\hbar^2}{\mu} \left( \frac{1}{n^2 a_0^2} - \frac{2}{a_0 \hat{r}} +\frac{n^2}{\hat{r}^2} \right), 
\end{equation}
hence
\begin{eqnarray} \label{BrBrdag1}
	\Big\langle n,n-1\Big|\hat{B}_{n-1}\hat{B}_{n-1}^{\dagger}\Big|n,n-1\Big\rangle = \frac{2\hbar^2}{\mu} \Bigg(& \frac{1}{n^2 a_0^2} - \frac{2}{a_0}\Big\langle n,n-1\Big|\frac{1}{\hat{r}}\Big|n,n-1\Big\rangle \nonumber \\
	& +n^2\Big\langle n,n-1\Big|\frac{1}{\hat{r}^2}\Big|n,n-1\Big\rangle \Bigg). 			
\end{eqnarray}
Now, we simply recall from the intertwining relation that
\begin{equation} \label{BrBrdag2}
	\hat{B}_{n-1}\hat{B}_{n-1}^{\dagger} =\hat{\mathcal H}_n-E_{n-1}= \hat{\cal{H}}_{n-1} + \frac{\hbar^2 n}{\mu \hat{r}^2}  - E_{n-1}.
\end{equation}
Substituting into Eq.~(\ref{BrBrdag1}), then yields
\begin{equation}
	\frac{\hbar^2n^2}{\mu}\Big\langle n,n-1\Big|\frac{1}{\hat{r}^{2}}\Big|n,n-1\Big\rangle = \frac{2\hbar^2}{\mu}\left( -\frac{1}{n^2 a_0^2} + n^2 \Big\langle n,n-1\Big|\frac{1}{\hat{r}^{2}}\Big|n,n-1\Big\rangle \right),
\end{equation}
after using the result for the first inverse moment in Eq.~(\ref{1/r}).
Hence we find
\begin{equation}
	\Big\langle n,n-1\Big|\frac{1}{\hat{r}^{2}}\Big|n,n-1\Big\rangle = \frac{1}{a_0^2 n^3 \left(n-\frac{1}{2}\right)},
\end{equation}
and combining this result with the one in Eq.~(\ref{r-2n}) we find
\begin{equation} \label{r-2res}
	\Big\langle n,l\Big|\frac{1}{\hat{r}^{2}}\Big|n,l\Big\rangle = \frac{1}{a_0^2 n^3 \left(l+\frac{1}{2}\right)}.
\end{equation}
Note that this derivation is completely algebraic and requires the discreteness of $l$ in carrying it out.

Now that we have determined the inverse second moment, we can find all additional inverse moments, bearing in mind that the wavefunction $\langle r|n,l\rangle$ behaves like $r^l$ as $r\to 0$. This means that the inverse moments exist up to $m=-2l-2$. Interestingly, the recurrence relation respects this result, in the sense that if one chooses an $m$ value that is too negative for a given $l$, the moment is indeterminate, because it has one factor in the denominator equal to zero.


We end this section with a short discussion of some exercises that may be assigned to the students learning this material. First, one can ask students to apply~the Kramers-Pasternack relation in Eq.~(\ref{Past2}) to derive moments for $m=3$ and $4$ and also $-3$ and $-4$ (higher order ones could also be assigned, but it rapidly becomes tedious to work out). This type of exercise gives students an opportunity to work with the recurrence relations and see how the formulas become increasingly complex for large $|m|$. One can also ask them to compute the standard deviation for the radial position operator using these relations.

Another useful problem is to have them derive  the inversion relation in Eq.~(\ref{Past1}) via induction. The base case with $m=0$ is established in Eq.~(\ref{r-2res}). We then assume that  it holds for all positive integers up to $m-1$. Next, we start with the left hand side of the equation and use the Kramers-Pasternack relation in Eq.~(\ref{Past2}) to relate it to lower values of $m$:
\begin{eqnarray}
   \langle n,l | \hat r^m|n,l\rangle&=\frac{(2m+1)n^2}{m+1}a_0\langle n,l|\hat r^{m-1}|n,l\rangle\nonumber\\
   &-\frac{m\left ((2l+1)^2-m^2\right )n^2}{4(m+1)}a_0^2\langle n,l|\hat r^{m-2}|n,l\rangle.
\end{eqnarray}
Then we use the inversion relation to relate each positive moment to a negative moment:
\begin{eqnarray}
   \langle n,l | \hat r^m|n,l\rangle&=\left (\frac{na_0}{2}\right )^{2m+1} \frac{(2l+m+1)!}{(2l-m)!}\frac{m}{m+1}\nonumber\\
   &\times \left (n^2a_0\frac{2m+1}{m} \langle n,l|{\hat r}^{-m-2}|n,l\rangle-\langle n,l|\hat r^{-m-1}|n,l\rangle\right ).
\end{eqnarray}
Finally, the Kramers-Pasternack relation (that involves the terms $-m-1$, $-m-2$ and $-m-3$) converts the sum of the expectation values of the two inverse moments to the expectation value of the inverse moment for $-m-3$. This gives
\begin{equation}
    \langle n,l | \hat r^m|n,l\rangle=\left (\frac{na_0}{2}\right )^{2m+3} \frac{(2l+m+2)!}{(2l-m-1)!}\langle n,l|r^{-m-3}|n,l\rangle.
\end{equation}
completing the induction [just shift $m\to m-1$ to determine the Pasternack inversion relation in Eq.~(\ref{Past1})].
This proof is a good opportunity to acquire a better understanding of both Pasternack relations.

One final problem that can be worked out is to examine similar recurrence relations for the isotropic simple harmonic oscillator in three dimensions~\cite{epstein_epstein}. Both the conventional recurrence relation and the inversion relation exist and can be established following a similar methodology as given here.

\section{Conclusions}

The Kramers-Pasternack relation (and to a lesser degree, the inversion relation) are often included in quantum mechanics instruction to varying degrees. We feel that they present an excellent opportunity to promote manipulations of operators and to develop skill in working with abstract  expressions for students learning quantum mechanics. The remarkable generality of these results also illustrates the power of working with operators. One of the challenges of working with these relations is that we need to be able to independently calculate the expectation values of the second inverse moment of $\hat r$. We showed how one can calculate this expectation value using only operator manipulations instead of performing a brute-force integration or using the Feynman-Hellman theorem. This new approach provides an alternative to the conventional approaches and has the potential of being easier to follow for students learning quantum mechanics.

\ack

This work was supported by the National Science Foundation under grant number PHY-1915130. In addition, JKF was supported by the McDevitt bequest at Georgetown University.

\end{document}